\pdfoutput=1

\documentclass[aps,twocolumn,amsmath,amssymb,preprintnumbers,floatfix]{revtex4-1}

\usepackage[utf8]{inputenc}
\usepackage{newtxtext}
\usepackage[upint]{newtxmath}
\usepackage{microtype}
\usepackage{textcomp}
\usepackage{eucal}
\usepackage{siunitx}
\usepackage{graphicx}
\usepackage{caption}
\usepackage{color}
\usepackage[caption=false]{subfig}
\usepackage{float}
\usepackage{ulem}

\usepackage{enumerate}
\usepackage{amsfonts}
\usepackage{amsmath}
\usepackage{amssymb}

\usepackage{bm}
\usepackage{color}
\usepackage{soul}

\usepackage{graphicx}

\usepackage[colorlinks,allcolors=blue]{hyperref}
\usepackage[capitalize]{cleveref}

\let\phi=\varphi
\let\epsilon=\varepsilon

\newcommand{\aang}{\text{ \AA}}

\definecolor{DarkRed}{rgb}{0.80,0,0}
\definecolor{lightBlue}{rgb}{0.5,0.63,0.9}
\definecolor{green}{rgb}{0,0.8,0.6}

\newcommand{\prlsection}[1]{\textit{#1}.\kern0.05em---\kern0.05em\ignorespaces}

\begin{document}
\title{RKKY interaction in a spin-split superconductor}
\author{Atousa Ghanbari}
\affiliation{\mbox{Center for Quantum Spintronics, Department of Physics, Norwegian University of Science and Technology,}\\NO-7491 Trondheim, Norway}

\author{Jacob Linder}
\affiliation{\mbox{Center for Quantum Spintronics, Department of Physics, Norwegian University of Science and Technology,}\\NO-7491 Trondheim, Norway}
\begin{abstract}
We determine theoretically the interaction between two magnetic impurities embedded in a spin-split $s$-wave superconductor. The spin-splitting in the superconductor gives rise to two different interaction types between the impurity spins, depending on whether their spins lie in the plane perpendicular to the spin-splitting field (Heisenberg) or not (Ising). For impurity separation distances exceeding $\xi_S$, we find that the magnitude of the spin-splitting can determine whether an antiferromagnetic or ferromagnetic alignment of the impurity spins is preferred by the RKKY interaction. Moreover, the Ising and Heisenberg terms of the RKKY interaction alternate on being the dominant term and their magnitudes oscillate as a function of distance between the impurities. 
\end{abstract}
\maketitle


\section{Introduction} 
\label{introduction}
\begin{figure}[H]
\subfloat{%
\includegraphics[width=1\columnwidth,trim= 0.01cm 3.6cm 0.01cm 3.6cm,clip=true]{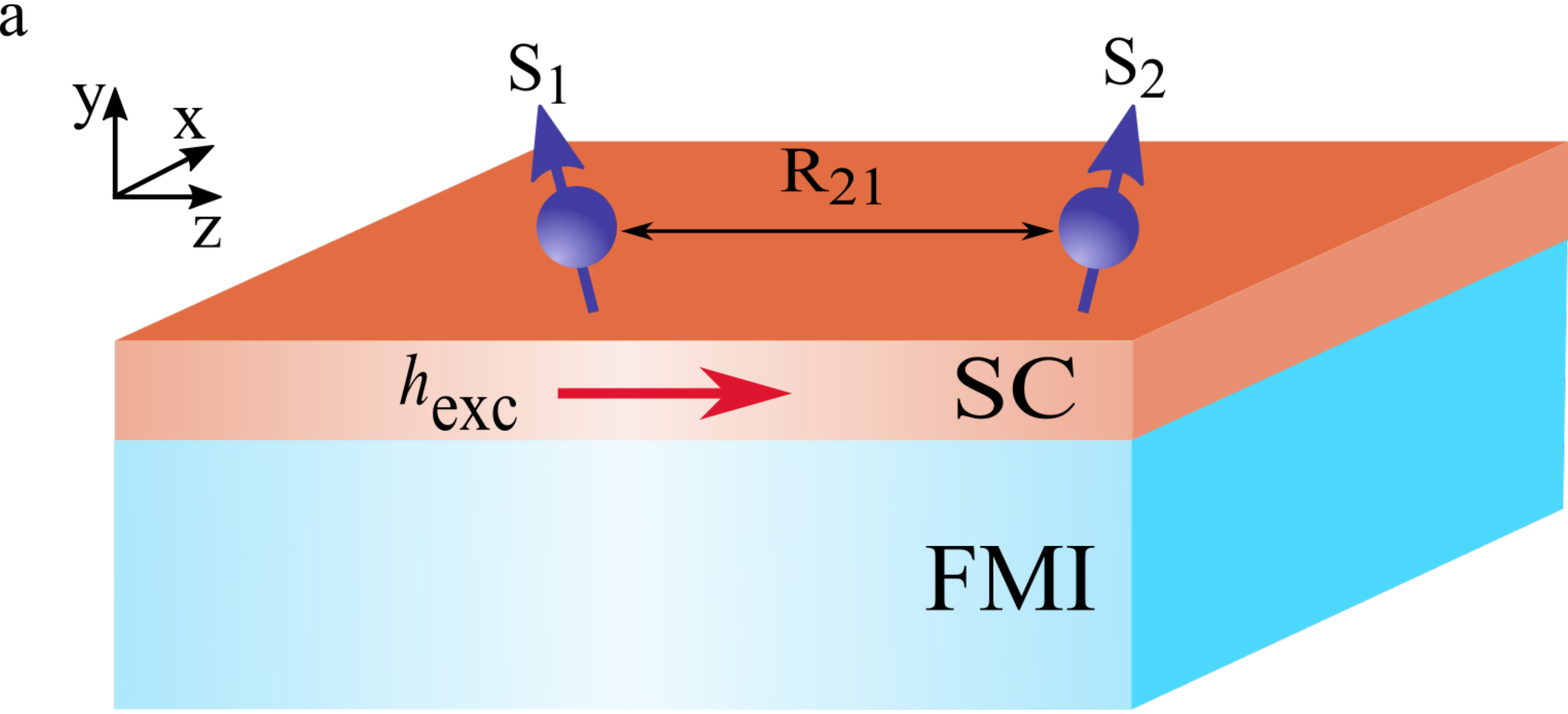}%
}
\vskip\baselineskip
\subfloat{%
\includegraphics[width=1\columnwidth,trim= 0.1cm 0.01cm 0.1cm 0.001cm,clip=true]{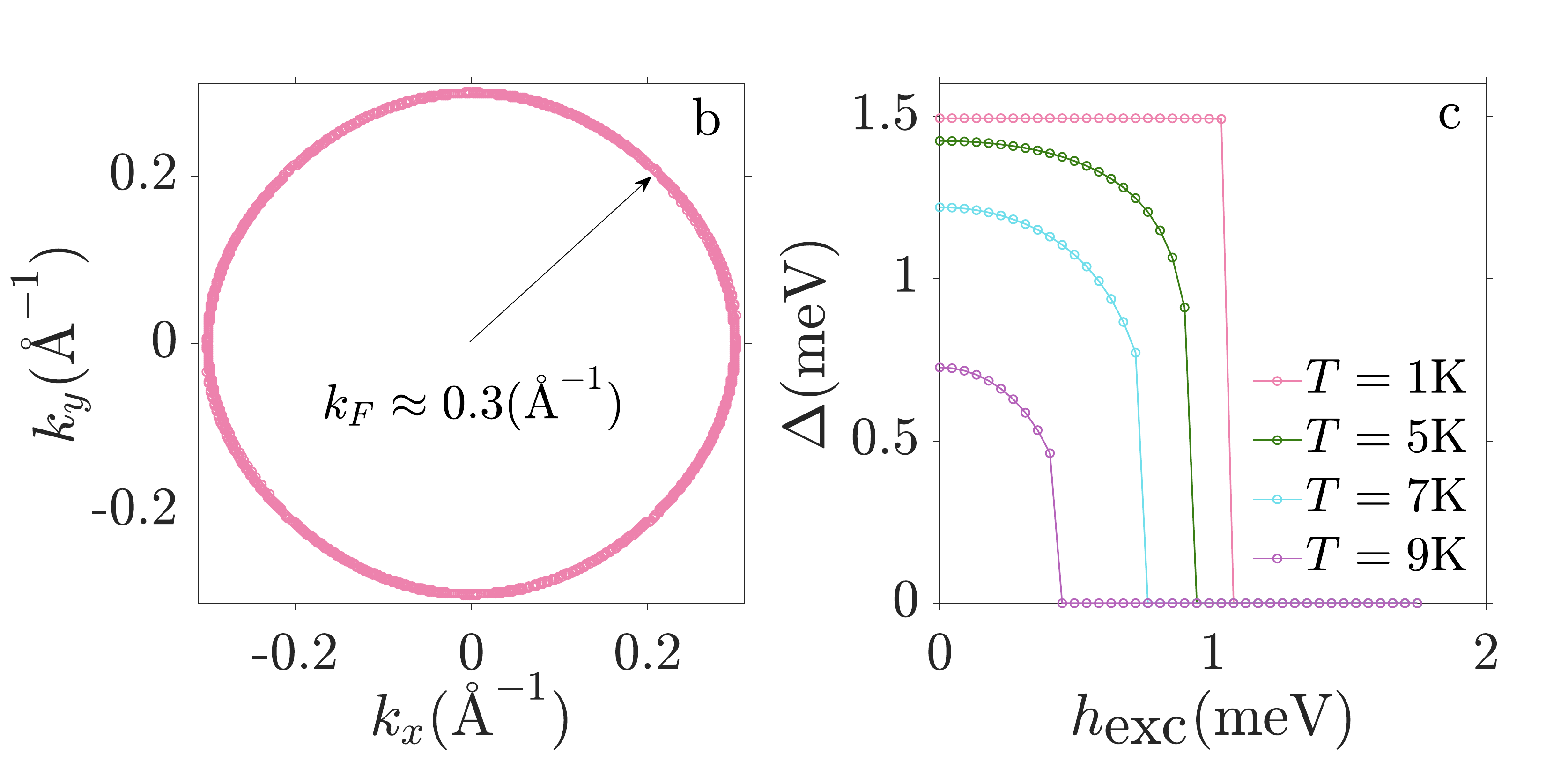}%
}
\caption{\small (a) Schematic illustration of a possible experimental realization of the system. A thin-film superconductor is placed on top of a ferromagnetic insulator. Quasiparticle reflection at the interface to the ferromagnetic insulator induces an effective spin-splitting field inside the superconductor. (b) Circular Fermi-surface with Fermi vector magnitude of $0.3\; \text{\AA}^{-1}$ used in our calculations. (c) Superconducting gap as a function of external exchange field for different temperature magnitudes.}
\label{fig:model}
\end{figure}

Superconductors have been experimentally demonstrated to exhibit strongly modified spin-dependent transport properties \cite{linder_natphys_15, eschrig_rpp_15} with respect to normal metals, such as spin relaxation times \cite{Spin_flip_relaxation_Sc_1, Spin_flip_relaxation_Sc_4, Spin_flip_relaxation_Sc_5,Spin_flip_relaxation_Sc_6} and magnetoresistance effects \cite{miao_prl_08}. Consequently, superconductors have the potential to advance research on spintronic devices, in which the spin of the electron is utilized as the information carrier instead of the electronic charge  \cite{Spintronics_1, Spintronics_2, Spintronics_3}. Intrinsically coexisting ferromagnetism and superconductivity, proposed more than 60 years ago \cite{ginzburg_jetp_57, matthias_prl_60, ferromagnetism_Sc_1964}, is only possible under rather strict conditions. On the other hand, by creating hybrid structures of ferromagnetic and superconducting materials, it is possible to study the interplay between these orders by virtue of the proximity effect
\cite{proximity}.\\
\indent The Ruderman–Kittel–Kasuya–Yosida (RKKY) interaction \cite{RKKKY_1,RKKY_2,RKKY_3} between magnetic impurities is an exchange interaction mediated by conduction electrons of the host material that the impurities are embedded in. This interaction has been vastly studied in different materials with spin-degeneracy, including systems with Dirac fermion excitations \cite{RKKY_Graphene_1, RKKY_Graphene_2, RKKY_TI_1} and superconducting materials \cite{alekseevskii_zetf_77, kochelaev_zetf_79, khusainov_zetf_96, aristov_zpb_97, dibernardo_natmat_19, ghanbari_scirep_21}. In a clean metal, the RKKY intercation decays as $R^{-D}$ where $R$ is the distance between the impurities and $D$ is the dimension of the system. Likewise, the interaction decays faster in higher dimensions also in superconducting systems.\\
\indent In the presence of spin-degeneracy, the RKKY interaction between magnetic impurities is isotropic in spin space and has no preferred direction for the impurity magnetic moments. On the other hand, it has been shown that in spin non-degenerate systems, the interaction can have different terms of the types Heisenberg, Ising and Dzyaloshinskii-Moriya (DM) \cite{DM}, depending on the spin structure of the host material. For instance, in a uniformly spin polarized system the Ising-term arises \cite{RKKY_Ising_1} whereas in systems with spin-orbit interactions a DM interaction term can emerge \cite{RKKY_DM_1, RKKY_DM_2,RKKY_DM_3,RKKY_in_Weyl_1,RKKY_in_Weyl_2}. In particular, the interaction between magnetic impurities located on top of an $s$-wave superconductor with Rashba spin-orbit coupling has been found to feature an additional DM term due to the spin-orbit coupling in the superconductor \cite{SOC_RKKY_s_wave_2015}. Similar results have been obtained for the interaction between magnetic impurities on top of a topological insulator with proximity-induced superconductivity from an $s$-wave superconductor \cite{SOC_RKKY_s_wave_with_TI_2014}.\\
\indent To the best of our knowledge, the RKKY interaction between magnetic impurities in a spin-split superconductor (see Fig. \ref{fig:model}) has not been studied. Such superconductors have in recent years been demonstrated to give rise to interesting spin-dependent thermoelectric effects and spin diffusion properties \cite{bergeret_rmp_17}. Due to the spin-splitting, the density of states in the superconductor acquires a large spin-dependent particle-hole asymmetry. Therefore, one might expect that the RKKY interaction could be modified compared to both the purely superconducting case and the case of a superconductor with spin-orbit interaction.\\ 
\indent In practice, a spin-split superconductor is achieved by either exposing a thin-film superconductor to a strong in-plane magnetic field or by growing a thin-film superconductor on top of a ferromagnetic insulator. 
In this case, the thickness of the  
\noindent superconductor has to be much smaller than the magnetic penetration depth $\lambda$. When the superconductor has a thickness smaller than the superconducting coherence length $\xi_S$, it can be well-approximated by a superconductor coexisting with a homogeneous spin-splitting field.\\ 
\indent In this paper, we will consider the RKKY interaction between two magnetic impurity atoms embedded in a spin-split conventional $s$-wave superconductor, contrasting it to the interaction between magnetic impurities in a normal metal subject to a spin-splitting field. While the RKKY interaction, in the normal metal case, is mediated by electrons, the RKKY in the superconducting case is mediated by quasiparticles that are a mix of electron and hole excitations. However, in both the superconducting and normal case a spin-splitting field induced via proximity to a ferromagnetic insulator lifts the spin degeneracy of the system. This causes the RKKY-interaction to have two parts: a Heisenberg- and Ising-term. In the present context, a Heisenberg term denotes the interaction energy obtained when the impurity spins lie in the plane perpendicular to the spin-splitting field. The Ising term describes the interaction for the case when the impurity spins are collinear with the spin-splitting field.

We find that it is possible to switch between an AFM and FM interaction between the magnetic impurities by adjusting the magnitude of the spin-splitting field. While this effect is in principle attainable even in the normal-state of the system, it is considerably more robust in the superconducting state where it occurs in a much larger regime of separation distances between the impurities compared to the normal-state. We discuss a possible experimental way to adjust the spin-splitting field strength in order to see this effect. Moreover, we find that the magnitudes of the Ising and Heisenberg terms of the RKKY interaction oscillate as a function of distance between the impurities, causing them to take turns on which is the dominant term. 

\indent This paper is structured as follows. We introduce the methodology used to compute the RKKY interaction in Sec.\! \ref{model}. In Sec.\! \ref{results}, we present a numerical evaluation of the expression for the RKKY interaction and discuss the underlying physics of its behavior. Finally, we summarize our findings in Sec.\! \ref{summary}.

\section{Model and Methods}
\label{model}

We consider a thin film s-wave superconductor in presence of a spin-splitting field which causes a spin-splitting in the electron bands, as shown in Fig.\! \ref{fig:model}. The superconductor is modelled by a tight-binding Hamiltonian including an attractive interaction between the electrons 

\begin{align} \label{Hamiltonian}
\begin{aligned}
    H_0 = &- \sum_{\langle i,j\rangle,\sigma} t_{ij} c_{i,\sigma}^{\dagger} c_{j,\sigma} + \sum_{i} V c_{i,\uparrow}^{\dagger} c_{i,\downarrow}^{\dagger} c_{i,\downarrow} c_{i,\uparrow}\\
    &-\sum_{i, \sigma} (\sigma h_\text{exc} + \mu) c_{i,\sigma}^{\dagger} c_{i,\sigma} .\\
\end{aligned}
\end{align}
The first term represents the nearest neighbour hopping term with $t_{ij} = t$ being the hopping parameter. The second term is the BCS on-site attractive interaction with $V < 0$ being the pairing strength. In the third term, $h_{\text{exc}}$ is the spin-splitting field. In our model, we consider this field to be oriented in the $z$-direction which is assumed to lie in the film plane of the superconductor. The Meissner response of the superconductor is well-known to be suppressed in a thin-film geometry when the field is applied in-plane and we may neglect orbital effects. 

\noindent We consider the system having continuous boundary conditions along both in-plane directions ($x$ and $z$ axes here). Using a Fourier transformation $c_{i \alpha}= \frac{1}{\sqrt{N}} \sum_{\bm{k}} e^{-i \bm{k} \cdot \bm{r}_i} c_{\bm{k}, \alpha}$ where $N$ is the total number of the lattice points, leads to the following form of the Hamiltonian in the $k$-space
\begin{equation} \label{eq:eq2}
H_0=\sum_{\bm{k},\sigma} (\zeta_{\bm{k}} - \sigma h_{\text{exc}}) c_{\bm{k},\sigma}^{\dagger} c_{\bm{k},\sigma} + \sum_{\bm{k} \bm{k^{\prime}}} V c_{\bm{k},\uparrow}^{\dagger} c_{-\bm{k},\downarrow}^{\dagger} c_{-\bm{k^{\prime}},\downarrow} c_{\bm{k^{\prime}},\uparrow},
\end{equation}
where $\zeta_k = -2t \big [ \cos(k_x a_x) + \cos(k_z a_z) ) \big ] - \mu $ and in it $a_x$($a_z$) is the lattice constant along $x$($z$) axis, also $\mu$ is the chemical potential. Here, we have redefined $V/N \to V$.\\ \indent Performing a mean-field treatment, we introduce the superconducting gap 
\begin{equation}\label{Gap}
\Delta = -V \sum_{\bm{k^{\prime}}} \langle c_{-\bm{k^{\prime}},\downarrow} c_{\bm{k^{\prime}},\uparrow} \rangle.    
\end{equation}
We then obtain,

\begin{align} \label{Hamiltonian_k_1}
\begin{aligned}
    H_0 = &\sum_{\bm{k},\sigma} (\zeta_{\bm{k}} - \sigma h_\text{exc}) c_{\bm{k},\sigma}^{\dagger} c_{\bm{k},\sigma} - \sum_{\bm{k},\sigma} \Delta c_{\bm{k},\uparrow}^{\dagger} c_{-\bm{k},\downarrow}^{\dagger}\\
    &- \sum_{\bm{k},\sigma} \Delta^{\ast} c_{-\bm{k},\downarrow} c_{\bm{k},\uparrow} - \frac{|\Delta|^2}{V}.\\
\end{aligned}
\end{align}
Using the following transformation (see Appendix \ref{Appendix_A} for details),

\begin{equation}\label{BdG_transformation}
   \left( {\begin{array}{c}
  c_{\bm{k} , \sigma}  \\
    c_{-\bm{k} , -\sigma}^{\dagger} 
    \end{array} } \right)= 
    \left( {\begin{array}{cc}
   \upsilon_{\bm{k}} & \sigma \nu_{\bm{k}}\\
   - \sigma \nu_{\bm{k}} & \upsilon_{\bm{k}}
    \end{array} } \right)
       \left( {\begin{array}{c}
  \gamma_{\bm{k} , \sigma}  \\
    \gamma_{-\bm{k} , -\sigma}^{\dagger} 
    \end{array} } \right),
\end{equation}
where,

\begin{equation}\label{eq:eq16}
\upsilon_{\bm{k}}= \frac{1}{\sqrt{2}} \sqrt{1+\frac{\zeta_{\bm{k}}}{\sqrt{\zeta_{\bm{k}}^2 + \Delta^2}}} , \nu_{\bm{k}}= \frac{1}{\sqrt{2}} \sqrt{1-\frac{\zeta_{\bm{k}}}{\sqrt{\zeta_{\bm{k}}^2 + \Delta^2}}},   
\end{equation}
the diagonalized form of $H_0$ will be
\begin{equation}\label{eq:eq9}
H_0= - \frac{|\Delta|^2}{V} + \sum_{\bm{k}} \zeta_{\bm{k}} - \sum_{\bm{k}} E_{\bm{k}} + \sum_{\bm{k},\sigma} E_{\bm{k}, \sigma}  \gamma_{\bm{k},\sigma}^{\dagger} \gamma_{\bm{k},\sigma} .      
\end{equation} 
Here, $E_{\bm{k}} = \sqrt{\zeta_{\bm{k}}^2 +  \Delta^2}$ and $E_{\bm{k}, \sigma}= E_{\bm{k}} - \sigma h_{\text{exc}}$. Expressing the electron operators in terms of the quasiparticle operators Eq.\! ~\eqref{BdG_transformation}, the gap equation takes the form 

\begin{align} \label{Gap_self_consistence}
    \begin{aligned}
        1 = & - \frac{V}{2} \sum_{\bm{k}} \frac{1}{2} \frac{1}{\sqrt{\zeta_{\bm{k}}^2 + \Delta^2}} \Big[\tanh(\frac{\beta}{2} (\sqrt{\zeta_{\bm{k}}^2 + \Delta^2} - h_{\text{exc}})) \\
        & + \tanh(\frac{\beta}{2} (\sqrt{\zeta_{\bm{k}}^2 + \Delta^2} + h_{\text{exc}})\Big].
    \end{aligned}
\end{align} 

In this study, the gap equation is solved self-consistently. Further, the free energy of the system is given by
 
\begin{align}\label{Free_Energy}
F = - \frac{|\Delta|^2}{V} + \sum_{\bm{k}} \zeta_{\bm{k}} - \sum_{\bm{k}} E_{\bm{k}} - \frac{1}{\beta}\sum_{\bm{k} , \sigma} \text{ln}(1 + e^{-\beta E_{\bm{k}, \sigma} }).
\end{align}

An important characteristic length scale in the system is the superconducting coherence length $\xi_S$ which is indicative of the size of the Cooper pairs. In the BCS formalism, this quantity for an isotropic $s$-wave superconductor is given by $\xi_S=\frac{\hbar v_F}{\pi \Delta_0}$, where $\hbar$ is the reduced Plank constant, $v_F$ is the Fermi velocity and $\Delta_0$ is the superconducting gap at zero temperature. The Fermi velocity is $v_F = \frac{1}{\hbar} \frac{d\zeta_k}{dk} |_{k=k_F}$.  

The main purpose of this paper is to determine the indirect exchange interaction between two magnetic impurity atoms mediated by the quasiparticles inside a superconductor described by the Hamiltonian in Eq.\!~\eqref{Hamiltonian}. The coupling between the quasiparticle spins and the magnetic impurities will be treated perturbatively. The total Hamiltonian can then be written as 
\begin{equation} \label{eq:eq1}
H=H_0+\Delta H,
\end{equation}
in which the first part is the non-perturbative Hamiltonian given by Eq.\! ~\eqref{Hamiltonian} and the second part is the perturbation defined by
\begin{equation}\label{eq:eq6}
\Delta H= J\sum_{j=1}^2  \bm{S}_j \cdot \bm{s}_j.
\end{equation}
Here, $J$ is the strength of the interaction between the spin of an impurity atom ($S_j$) and an itinerant spin ($s_j$) at lattice site $j$. The impurity spin is treated classically like a normal vector and itinerant spin is treated quantum mechanically and represented by the operator $s_j = \sum_{\alpha \beta} \bm{\sigma}_{\alpha \beta} c_{j \alpha}^{\dagger} c_{j \beta}$. Here, $\boldsymbol{\sigma} = (\sigma_x, \sigma_y, \sigma_z)$ is the Pauli matrix vector. Performing a Fourier transformation, the perturbation term in the Hamiltonian becomes

\begin{equation}\label{Delta_H}
\Delta H=\sum_{\bm{k},\bm{k^{\prime}} \atop \alpha, \beta} \sum_j \frac{J}{N} e^{i(\bm{k}-\bm{k^{\prime}}) \cdot \bm{r}_j} (\bm{S}_j \cdot \bm{\sigma}_{\alpha \beta}) c_{\bm{k},\alpha}^{\dagger} c_{\bm{k^{\prime}},\beta}.
\end{equation}

By means of Eq.\! \eqref{BdG_transformation}, we change the $c_{\bm{k},\alpha}$ operators into quasiparticle operators. Then, by means of a Schrieffer-Wolff transformation (SWT), the effective interaction between the magnetic impurity atoms is obtained to second order in the coupling $J$. To obtain the effective interaction, we consider a unitary matrix $U$ of the form $U=e^{iS}$. The unitary transformation of the total Hamiltonian $H$ is then, 

\begin{equation}\label{Unitary_transformation}
\tilde{H}= U H U^{\dagger} = e^{iS} H e^{-iS}.    
\end{equation}
\noindent The above equation may be expanded as

\begin{align}\label{eq:eq32}
    \begin{aligned}
        \tilde{H} = H_0 + \Delta H + i[S,H_0] + i[S,\Delta H] - \frac{1}{2}[S,[S,H_0]] + O(J^3),
    \end{aligned}
\end{align}

\noindent where we take $S= J S^{\prime}$ and discard higher order terms in $J$. This leads to the following effective Hamiltonian for the system,

\begin{equation}\label{eq:eq33}
\tilde{H}= H_0 + \Delta H + i[S,H_0] + i[S,\Delta H] - \frac{1}{2}[S,[S,H_0]].    
\end{equation}

\noindent We now choose the unitary transformation $S$ so that $\Delta H + i[S,H_0] = 0$ and the effective Hamiltonian becomes $\tilde{H} = H_0 + \frac{i}{2}[S,\Delta H]$. In order to accomplish this, we consider the following Ansatz for $S$ 

\begin{equation}\label{unitary_matrix}
\begin{aligned}
S= & \sum_{\bm{k},\bm{k^{\prime}} \atop \alpha, \beta} (A_{\bm{k},\bm{k^{\prime}}        \atop \alpha, \beta} \gamma_{\bm{k} , \alpha}^{\dagger} \gamma_{\bm{k^{\prime}}      , \beta} + B_{\bm{k},\bm{k^{\prime}} \atop \alpha, \beta} \gamma_{\bm{k} ,          \alpha}^{\dagger} \gamma_{-\bm{k^{\prime}} , -\beta}^{\dagger} \\
    &+ C_{\bm{k},\bm{k^{\prime}} \atop \alpha, \beta} \gamma_{-\bm{k} , -\alpha} \gamma_{\bm{k^{\prime}} , \beta} + D_{\bm{k},\bm{k^{\prime}} \atop \alpha, \beta} \gamma_{-\bm{k} , -\alpha} \gamma_{-\bm{k^{\prime}} , -\beta}^{\dagger}). 
\end{aligned}
\end{equation}
\noindent  Computing the commutator $[S,H_0]$, and requiring $\Delta H + i[S,H_0]=0$, the coefficients in $S$ are found to be

\begin{align}\label{coeficients}
\begin{aligned}
A_{\bm{k},\bm{k^{\prime}} \atop \alpha, \beta}= & i\sum_j \frac{J}{N} e^{i(\bm{k}-\bm{k^{\prime}}) \cdot \bm{r}_j} (\bm{S}_j \cdot \bm{\sigma}_{\alpha \beta}) \frac{\upsilon_{\bm{k}}^{\ast} \upsilon_{\bm{k^{\prime}}}}{E_{\bm{k^{\prime}} , \beta} - E_{\bm{k} , \alpha}},  \\  
B_{\bm{k},\bm{k^{\prime}} \atop \alpha, \beta}=  & - \beta i\sum_j \frac{J}{N} e^{i(\bm{k}-\bm{k^{\prime}}) \cdot \bm{r}_j} (\bm{S}_j \cdot \bm{\sigma}_{\alpha \beta}) \frac{\upsilon_{\bm{k}}^{\ast} \nu_{\bm{k^{\prime}}}}{E_{-\bm{k^{\prime}},-\beta} + E_{\bm{k}, \alpha}}, \\   
C_{\bm{k},\bm{k^{\prime}} \atop \alpha, \beta}=   & \alpha i\sum_j \frac{J}{N} e^{i(\bm{k}-\bm{k^{\prime}}) \cdot \bm{r}_j} (\bm{S}_j \cdot \bm{\sigma}_{\alpha \beta}) \frac{\nu_{\bm{k}}^{\ast} \upsilon_{\bm{k^{\prime}}}}{E_{\bm{k^{\prime}}, \beta} + E_{-\bm{k}, -\alpha}}, \\    
D_{\bm{k},\bm{k^{\prime}} \atop \alpha, \beta}= & \alpha \beta i\sum_j \frac{J}{N} e^{i(\bm{k}-\bm{k^{\prime}}) \cdot \bm{r}_j} (\bm{S}_j \cdot \bm{\sigma}_{\alpha \beta}) \frac{\nu_{\bm{k}}^{\ast} \nu_{\bm{k^{\prime}}}}{-E_{-\bm{k^{\prime}}, -\beta} + E_{-\bm{k}, -\alpha}}.    
\end{aligned}
\end{align}

The final form of the effective Hamiltonian $\tilde{H}$ is obtained after calculating $[S,\Delta H]$. In this Hamiltonian, we neglect terms representing feedback from the impurity spin on the superconductor. Feedback from the impurities would ideally be included by self-consistently taking into account both the effect of the presence of the superconductor on the impurity spins and the effect of the impurity spins on the superconducting gap, giving rise to spatial variation of the superconducting order parameter. As the density of impurities in the system is very low, neglecting feedback from the impurities can be justified. 

Computing the expectation value of the effective Hamiltonian $\tilde{H}$ (given explicitly in Appendix ~\ref{Apendix_B}) leads to two different terms in the interaction energy between the two magnetic impurities: a 2D Heisenberg-like ($E_H$) and Ising-like ($E_I$) interaction

\begin{align}\label{eq:e0}
\langle \tilde{H}\rangle = E_0 +  2 E_I (S_1^zS_2^z) + 2 E_H (S_1^x S_2^x + S_1^y S_2^y),
\end{align}
where $E_0$ is a constant. In the following section \ref{results}, we will consider these $E_I$ and $E_H$ terms in more detail analytically and then evaluate them numerically to determine the nature of the RKKY interaction in a spin-split superconductor.

\section{Results and discussion}
\label{results}

\subsection{Analytical}

The physical significance of the RKKY interaction terms $E_I$ and $E_H$ is described as follows. The Ising term $E_I$ determines the strength of the interaction between the magnetic impurities when they are oriented collinearly to the spin-splitting field. For $E_I>0$, the interaction prefers an AFM alignment of the impurity spins. For $E_I<0$, they prefer a FM alignment. The Heisenberg term $E_H$ determines the strength of the interaction between the magnetic impurities when they lie in the plane perpendicular to the spin-splitting field. The same considerations regarding the sign for $E_H$ hold as for the Ising term.

The explicit expression for the RKKY Ising-like interaction between the spin of impurity atom 1 and the spin of impurity atom 2 is found to be

\begin{align} \label{E_I}
    \begin{aligned}
        & E_{I}=  -\frac{1}{2} \sum_{\bm{k},\bm{k^{\prime}} } \Big(\frac{J}{N}\Big)^2 e^{i (\bm{k^{\prime}} - \bm{k} ) \cdot \bm{R_{21}} } \Big[ (|\upsilon_{\bm{k}} \upsilon_{\bm{k^{\prime}}}|^2 + | \nu_{\bm{k}} \nu_{\bm{k^{\prime}}}|^2) \\
        &\times (\frac{n(E_{\bm{k}, \uparrow})-n(E_{\bm{k^{\prime}} , \uparrow})}{E_{\bm{k^{\prime}} , \uparrow}-E_{\bm{k}, \uparrow}} + \frac{n(E_{\bm{k}, \downarrow})-n(E_{\bm{k^{\prime}} , \downarrow})}{E_{\bm{k^{\prime}} , \downarrow}-E_{\bm{k}, \downarrow}} )  -2 \upsilon_{\bm{k}}^{\ast} \upsilon_{\bm{k^{\prime}}} \nu_{\bm{k}}^{\ast} \nu_{\bm{k^{\prime}}} \\
        &\times (\frac{n(E_{\bm{k^{\prime}}, \uparrow})-n(E_{\bm{k} , \uparrow})}{E_{\bm{k^{\prime}} , \uparrow}-E_{\bm{k}, \uparrow}} + \frac{n(E_{\bm{k^{\prime}}, \downarrow})-n(E_{\bm{k} , \downarrow})}{E_{\bm{k^{\prime}} , \downarrow}-E_{\bm{k}, \downarrow}})  -2 \upsilon_{\bm{k}}^{\ast} \upsilon_{\bm{k^{\prime}}} \nu_{\bm{k}}^{\ast} \nu_{\bm{k^{\prime}}} \\
        &\times (  \frac{1-n(E_{\bm{k}, \uparrow})-n(E_{\bm{k^{\prime}} , \downarrow})}{E_{\bm{k^{\prime}} , \downarrow}+E_{\bm{k}, \uparrow}} + \frac{1-n(E_{\bm{k}, \downarrow})-n(E_{\bm{k^{\prime}} , \uparrow})}{E_{\bm{k^{\prime}} , \uparrow}+E_{\bm{k}, \downarrow}})  
        \\
        & + ( \frac{1-n(E_{\bm{k}, \uparrow})-n(E_{\bm{k^{\prime}} , \downarrow})}{E_{\bm{k} , \uparrow}+E_{\bm{k^{\prime}}, \downarrow}} +  \frac{1-n(E_{\bm{k}, \downarrow})-n(E_{\bm{k^{\prime}} , \uparrow})}{E_{\bm{k} , \downarrow}+E_{\bm{k^{\prime}}, \uparrow}} ) \\ 
        & \times 2 \upsilon_{\bm{k}}^{\ast} \upsilon_{\bm{k}} \nu_{\bm{k^{\prime}}}^{\ast} \nu_{\bm{k^{\prime}}} \Big]. 
    \end{aligned}
\end{align}

\noindent Here, $\bm{R_{21}} = \bm{r_2} - \bm{r_1}$ is the relative distance between the two impurity atoms and $n(E_{\bm{k}, \sigma}) = (1+e^{\beta E_{\bm{k}, \sigma}})^{-1}$ is the Fermi-Dirac distribution function. The Heisenberg-like term in the RKKY interaction energy is

\begin{align}\label{E_H}
    \begin{aligned}
        & E_{H} = -\frac{1}{2} \sum_{\bm{k},\bm{k^{\prime}} } \Big(\frac{J}{N}\Big)^2 e^{i (\bm{k^{\prime}} - \bm{k} ) \cdot \bm{R_{21}}} \Big[(|\upsilon_{\bm{k}} \upsilon_{\bm{k^{\prime}}}|^2 + | \nu_{\bm{k^{\prime}}} \nu_{\bm{k}}|^2) \\
        & \times ( \frac{n(E_{\bm{k}, \uparrow})-n(E_{\bm{k^{\prime}} , \downarrow})}{E_{\bm{k^{\prime}} , \downarrow}-E_{\bm{k}, \uparrow}} + \frac{n(E_{\bm{k}, \downarrow})-n(E_{\bm{k^{\prime}} , \uparrow})}{E_{\bm{k^{\prime}} , \uparrow}-E_{\bm{k}, \downarrow}} ) -2 \upsilon_{\bm{k}}^{\ast} \upsilon_{\bm{k^{\prime}}} \nu_{\bm{k}}^{\ast} \nu_{\bm{k^{\prime}}} \\
        & \times ( \frac{n(E_{\bm{k^{\prime}}, \downarrow})-n(E_{\bm{k} , \uparrow})}{E_{\bm{k^{\prime}} , \downarrow}-E_{\bm{k}, \uparrow}} +   \frac{n(E_{\bm{k^{\prime}}, \uparrow})-n(E_{\bm{k} , \downarrow})}{E_{\bm{k^{\prime}} , \uparrow}-E_{\bm{k}, \downarrow}}) -2 \upsilon_{\bm{k}}^{\ast} \upsilon_{\bm{k^{\prime}}} \nu_{\bm{k}}^{\ast} \nu_{\bm{k^{\prime}}} \\  
        & \times (\frac{1-n(E_{\bm{k}, \uparrow})-n(E_{\bm{k^{\prime}} , \uparrow})}{E_{\bm{k^{\prime}} , \uparrow}+E_{\bm{k}, \uparrow}} + \frac{1-n(E_{\bm{k}, \downarrow})-n(E_{\bm{k^{\prime}} , \downarrow})}{E_{\bm{k^{\prime}} , \downarrow} + E_{\bm{k}, \downarrow}} ) \\
        & + (\frac{1-n(E_{\bm{k}, \uparrow})-n(E_{\bm{k^{\prime}} , \uparrow})}{E_{\bm{k} , \uparrow}+E_{\bm{k^{\prime}}, \uparrow}} + \frac{1-n(E_{\bm{k}, \downarrow})-n(E_{\bm{k^{\prime}} , \downarrow})}{E_{\bm{k} , \downarrow}+E_{\bm{k^{\prime}}, \downarrow}}) \\
        & \times 2 \upsilon_{\bm{k}}^{\ast} \upsilon_{\bm{k}} \nu_{\bm{k^{\prime}}}^{\ast} \nu_{\bm{k^{\prime}}} \Big]. 
    \end{aligned}
\end{align}

In the limiting case of $h_\text{exc} = 0$, the two above terms are equal. The system then displays a normal 3D Heisenberg-like interaction between the two impurity atoms hosted by an $s$-wave superconductor, which is spin isotropic as it should.

\subsection{Numerical}
Proceeding to a numerical evaluation of $E_H$ and $E_I$, we consider a system of $N = 800 \times 800$ lattice points in the $xz$ plane. We choose $V$ so that the zero-temperature superconducting gap takes the value $\Delta \approx 1.5$ meV.  
The lattice constants are set to $a_x = a_z = 3.5 \aang$. The hopping parameter and chemical potential magnitudes are taken to be $t = 0.2$ eV and $\mu = -0.6$ eV, respectively. The chemical potential is chosen to provide us with a circular Fermi surface as shown in Fig.\! \ref{fig:model}\! (b). The superconducting gap at $T = 0 K$, the Fermi velocity, the Fermi wave vector, and coherence length take the values $\Delta_0 = 1.49$ meV, $v_F = 1.91 \times 10^{5} \frac{\text{m}}{\text{s}}$, $k_F \approx 0.3 \aang$ and $\xi_S = 269 \aang$, respectively. \\ 
\indent Fig.\! \ref{fig:model}\! (c) illustrates the gap versus the spin-splitting field for different temperatures. A nontrivial solution to the gap equation does not guarantee that the superconducting phase is the ground state of the system. For each temperature and field strength, the ground state of the system (either $\Delta = 0$ or $\Delta \neq 0$) has therefore been determined by computing the free energy of the system given in Eq.\! \eqref{Free_Energy}. At $\text{T} \approx 0$ K the largest spin-splitting which allows for a superconducting phase as the ground-state is approximately $h_{\text{exc}} \approx 0.7 \Delta_0$ which is around $1.07 \text{meV}$ with our set of parameters. This is consistent with the Clogston-Chandrasekhar limit. It is also seen from the figure that increasing temperature reduces the gap until a phase transition occurs at the critical temperature which is around $T_C = 9.829 \text{K}$ for $h_\text{exc}=0$. A superconductor with a similar set of parameters as chosen above is niobium (Nb) with a critical temperature $T_C \approx 9.2 \text{K}$ \cite{niobium_T_C}.    

\subsubsection{Low temperatures $T\ll T_c$}

\begin{figure}[H]
\centering
\subfloat{%
\includegraphics[width=1\columnwidth,trim= 0.2cm 0.01cm 1.0cm 0.01cm,clip=true]{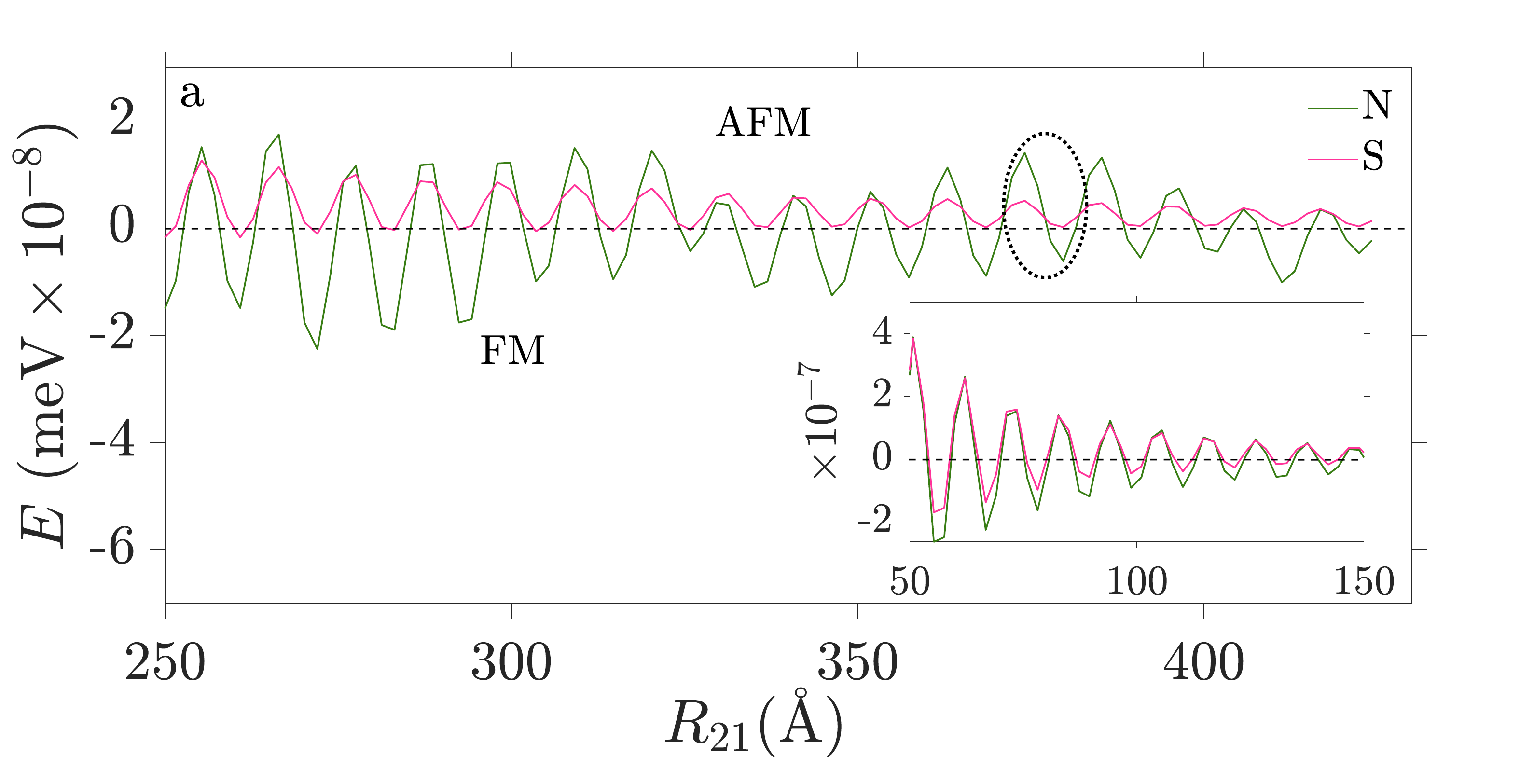}%
}
\vskip\baselineskip
\subfloat{%
\includegraphics[width=1.05\columnwidth,trim= 0.2cm 0.01cm 0.01cm 0.01cm,clip=true]{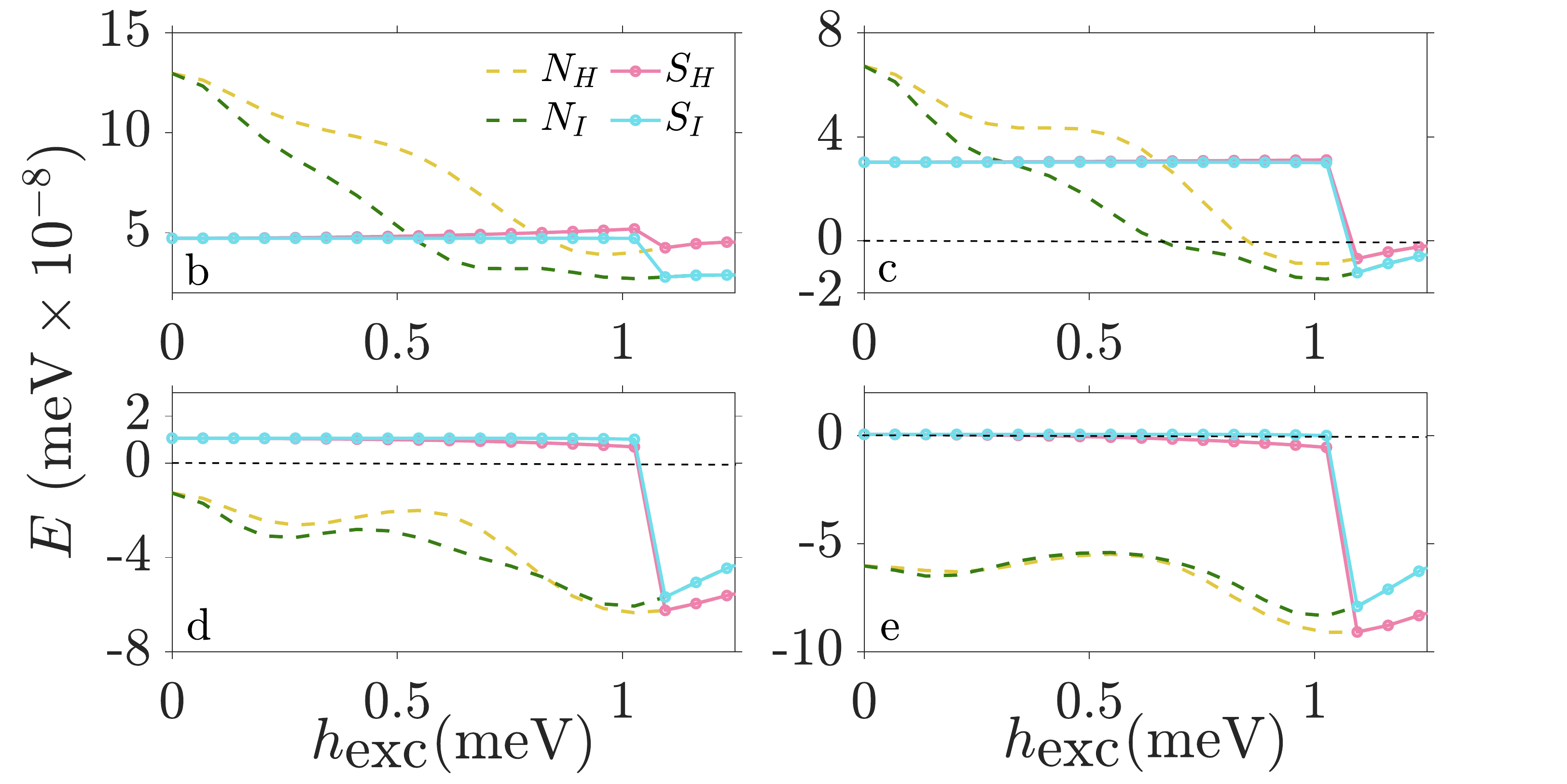}%
}
\caption{\small (a) RKKY energy versus $R_{21}$ when $h_{\text{exc}} = 0$. The inset represents the energies for distances smaller than coherence length. Furthermore, the energies as a function of exchange field for (b) $R_{21} = 374.7 \text{\AA}$ (c) $R_{21} = 376.18 \text{\AA}$ (d) $R_{21} = 377.59 \text{\AA}$ (e) $R_{21} = 379 \text{\AA}$ are computed. Here, $N_H (N_I)$ is the Heisenberg (Ising) RKKY interaction energy for the normal-state of the system while $S_H (S_I)$ is Heisenberg (Ising) RKKY interaction energy for the superconducting phase. The temperature is fixed at $1$ k.}
\label{fig:T_1K_vs_R21_and_h}
\end{figure}

We start by considering temperatures well within the superconducting phase, $T\ll T_c$, and here set $T = 1$ K. The strength of the exchange interaction between the impurity spins and the quasiparticle spins is taken to be $J = 1$ \text{meV}. For $h_\text{exc} = 0$, the RKKY energies Eq.\! \eqref{E_I} and Eq.\! \eqref{E_H} are presented as a function of the distance between the two impurity atoms in Fig.\! \ref{fig:T_1K_vs_R21_and_h} (a). The RKKY energy goes to zero as $R_{21}$ increases as seen in the inset of Fig.\! \ref{fig:T_1K_vs_R21_and_h} (a). The effect of the superconducting gap is primarily to shift the RKKY energy above zero for distances larger than coherence length $\xi_S$. Consequently, the interaction prefers an AFM orientation of the impurity spins at such distances. In the normal-state of the system, the RKKY signal changes sign between FM and AFM alignment, also for large distances. These results are consistent with previous literature.

Considering instead the case where the spin-splitting field $h_\text{exc}$ is present, an interesting possibility with regard to the tunability of the RKKY interaction opens up. Since the RKKY interaction $E$ is positive in the superconducting state at $h_\text{exc}=0$ for $R_{21} > \xi_S$ whereas it oscillates in the normal-state, driving the system through a phase transition by increasing $h_\text{exc}$ above its critical value will change the sign of the RKKY interaction whenever the oscillations in the normal-state causes $E<0$. We illustrate this in Fig.\! \ref{fig:T_1K_vs_R21_and_h} (b)-(e) which shows the RKKY energies at four different separation distances taken from the dashed oval region marked in Fig.\! \ref{fig:T_1K_vs_R21_and_h} (a). 

\begin{figure}[H]
\centering
\vskip\baselineskip
\subfloat{%
\includegraphics[width=1\columnwidth,trim= 0.2cm 0.01cm 0.01cm 0.01cm,clip=true]{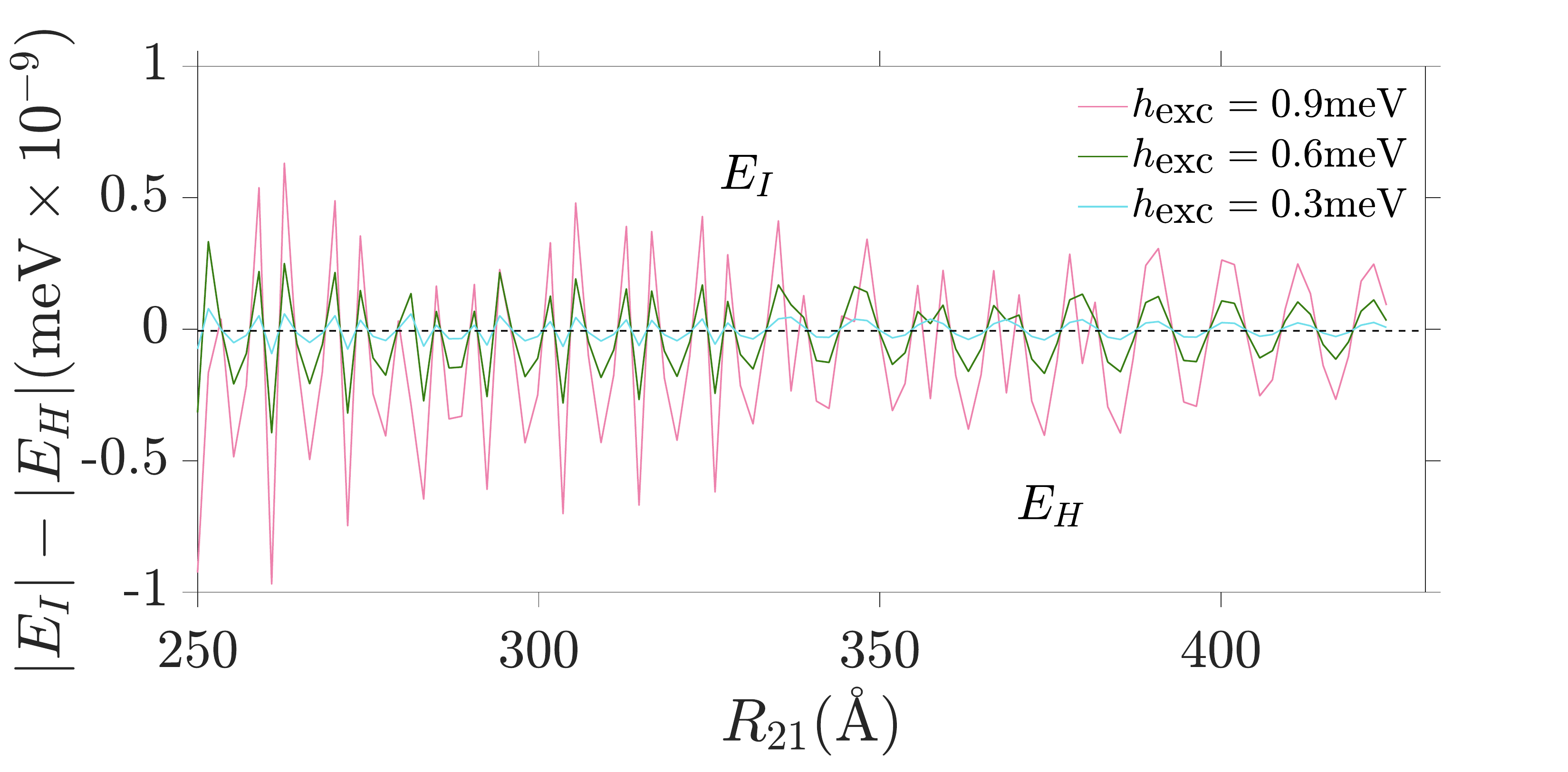}%
}
\caption{\small Difference between the Ising and Heisenberg energies for three different values of the spin-splitting field at $T = 1$ K for the superconducting state. }
\label{fig:T_1K_vs_R21_E_I_minus_E_H}
\end{figure}

It can be seen from Fig.\! \ref{fig:T_1K_vs_R21_and_h} (c)-(e) that by increasing $h_\text{exc}$ one can change the RKKY energy sign from AFM alignment into FM alignment and vice versa. In contrast to the normal-state of the system where $E$ varies significantly with $h_\text{exc}$, the RKKY interaction in the superconducting phase is practically independent of $h_\text{exc}$ in comparison. This can be understood from the fact that the superconducting gap changes very slowly as a function of $h_{\text{exc}}$ for low temperatures, as seen in Fig.\! \ref{fig:model}\! (c). As a result, an abrupt change occurs once the phase transition to the normal-state takes place, which can cause a sign change
\noindent in the RKKY interaction. A sign change can in principle also occur in the normal-state of the system, as shown in Fig.\! \ref{fig:T_1K_vs_R21_and_h}\! (c), but this effect is far less robust than the one observed in the superconducting state. In the normal-state of the system, the sign-change can only occur at carefully chosen separation distances $R_{21}$, whereas the sign-change occurs in the superconducting state for a much larger set of separation distances. More precisely, when the separation distance between the impurities is larger than the coherence length, the sign-change occurs in the superconducting state whenever the normal-state RKKY oscillations cause $E$ to be negative. In principle, above the coherence length, this corresponds to half of all separation distances.\\
\indent It is also of interest to determine whether the interaction between the magnetic impurities in the system favor their spins being collinear with the spin-splitting field or lying in the plane perpendicular to it. To this end, we compute the difference between the magnitude of the Ising and Heisenberg energies ($|E_I| - |E_H|$) as a function of distance between the impurities for several different values of the spin-splitting field in the superconducting phase (Fig.\! \ref{fig:T_1K_vs_R21_E_I_minus_E_H}). The term which is largest in magnitude will dictate whether the interaction prefers the impurity spins to orient in the plane normal to the exchange field or collinearly with it. The sign of the largest term thereafter 
\noindent determines whether the interaction prefers the impurity spins to orient parallell or antiparallell. The difference in magnitude between the Ising and Heisenberg interaction energies oscillates as a function of separation distance, making the two interaction terms take turns on being dominant.

\subsubsection{High temperatures $T \lesssim T_c$}

\begin{figure}[H]
\subfloat{%
\includegraphics[width=1\columnwidth,trim= 0.2cm 0.01cm 1.0cm 0.01cm,clip=true]{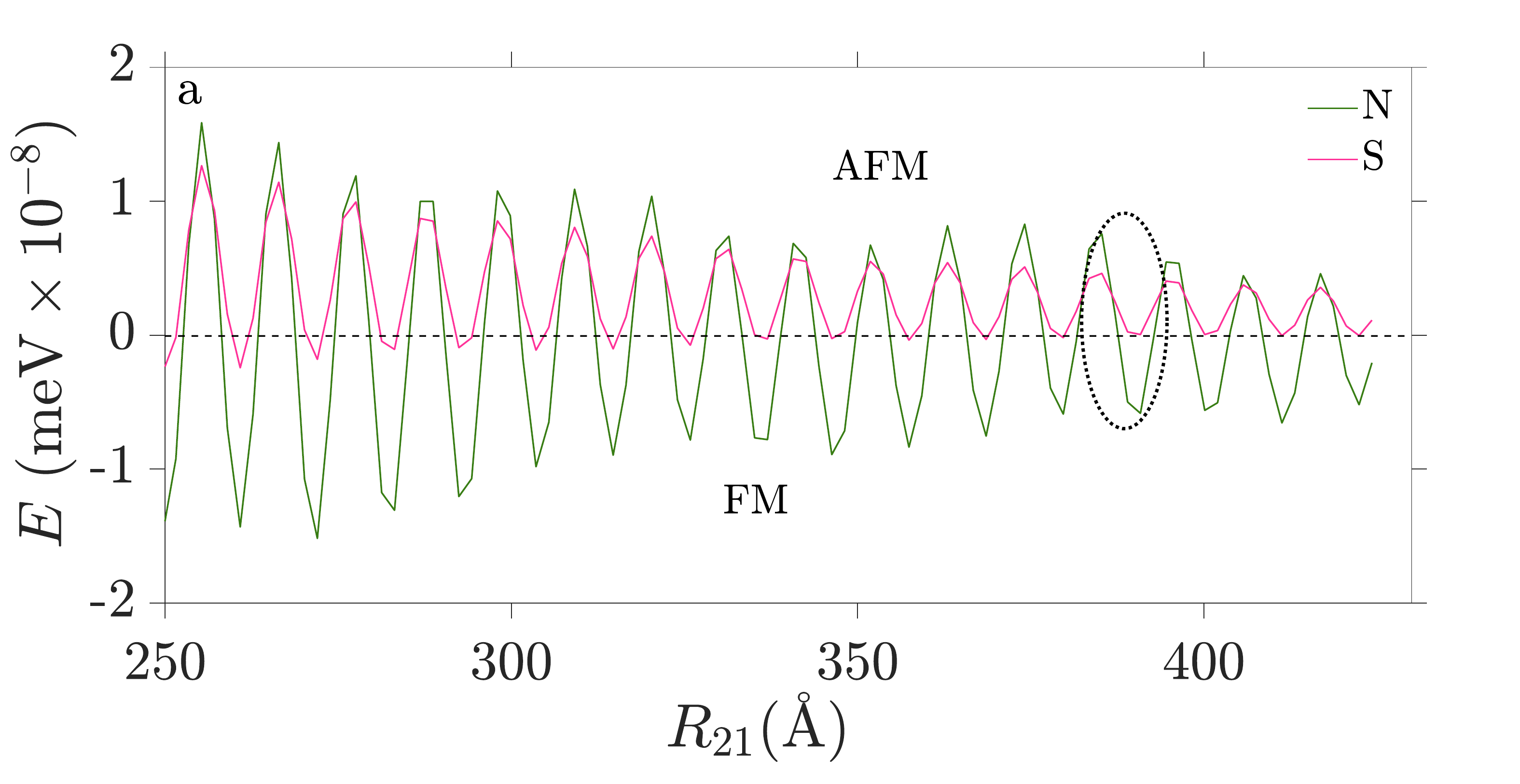}%
}
\vskip\baselineskip
\subfloat{%
\includegraphics[width=1.05\columnwidth,trim= 0.2cm 0.01cm 0.01cm 0.01cm,clip=true]{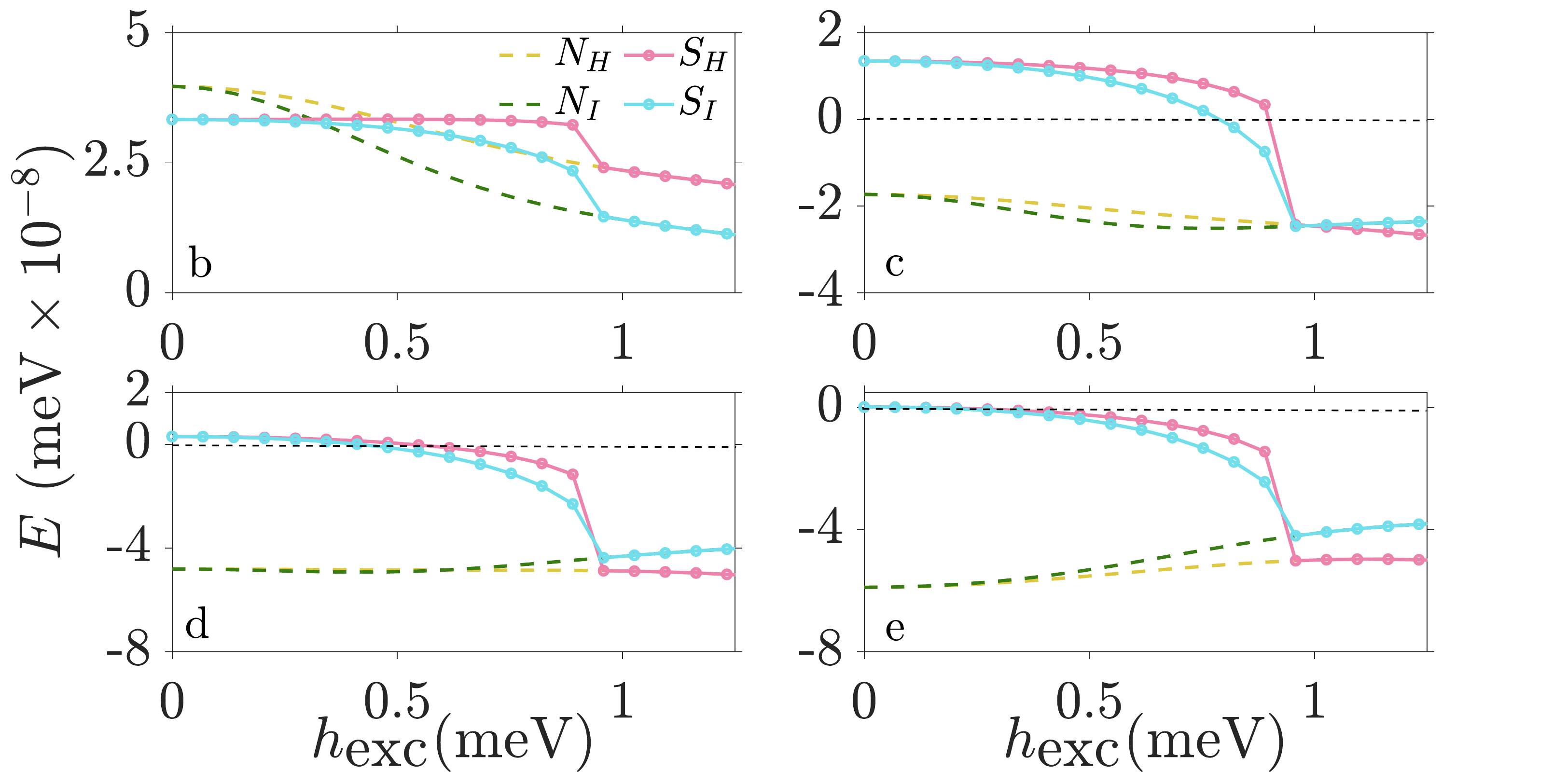}%
}
\caption{\small (a) RKKY energy versus $R_{21}$ when $h_{\text{exc}} = 0$. The RKKY energies as a function of exchange field for (b) $R_{21} = 386.561 \text{\AA}$ (c) $R_{21} = 387.975 \text{\AA}$ (d) $R_{21} = 388.908 \text{\AA}$ (e) $R_{21} = 390.803 \text{\AA}$ are computed. Here, $N_H (N_I)$ is Heisenberg (Ising) RKKY interaction energy for normal metal state and $S_H (S_I)$ is Heisenberg (Ising) RKKY interaction energy for the superconducting phase. The temperature is fixed at $4$ K.}
\label{fig:T_4K_vs_R21_and_h}
\end{figure}

In order to show the effect of temperature on the results, we consider in this section $T = 4$ K, taken to represent the regime $T \lesssim T_c$. Similarly to the previous section, we first compute the change in the RKKY energy as a function of $R_{21}$ when no spin-splitting field is present for both the normal-state and superconducting phase of the system in Fig.\! \ref{fig:T_4K_vs_R21_and_h} (a). The results are qualitatively similar to the low-temperature case. For $R_{21} \ll \xi_S$, the signal oscillates both in the normal and superconducting state, while above $\xi_S$ the interaction between the magnetic impurities is AFM in the superconducting state.\\ 
\indent When the spin-splitting field is present, as shown in Figs.\! \ref{fig:T_4K_vs_R21_and_h} (b)-(e), the RKKY interaction in the superconducting state is more strongly affected by a change in $h_\text{exc}$ than in the low-temperature case considered in the previous section. This can be understood from the exchange field having a larger effect on the superconducting order parameter at higher temperatures, as displayed in Fig.\! \ref{fig:model}(c). As a result, it becomes easier to change the sign of the RKKY interaction energies $E_I$ and $E_H$ by increasing $h_\text{exc}$ while still remaining in the superconducting phase of the system. In fact, it can be seen from Figs.\! \ref{fig:T_4K_vs_R21_and_h} (c)-(e) that the sign change can occur for much lower spin-splitting fields than in the low-temperature case. We also find that a sign-change of the RKKY interaction becomes more difficult to achieve in the normal-state of the system and no such sign-change is observed in any of the plots in Fig.\! \ref{fig:T_4K_vs_R21_and_h}. In fact, the sign-change now only occurs at highly selective separation distances $R_{21}$ in the normal-state where the RKKY-oscillations cause the interaction to almost vanish.

\begin{figure}[H]
\vskip\baselineskip
\subfloat{%
\centering
\includegraphics[width=1\columnwidth,trim= 0.2cm 0.01cm 0.01cm 0.01cm,clip=true]{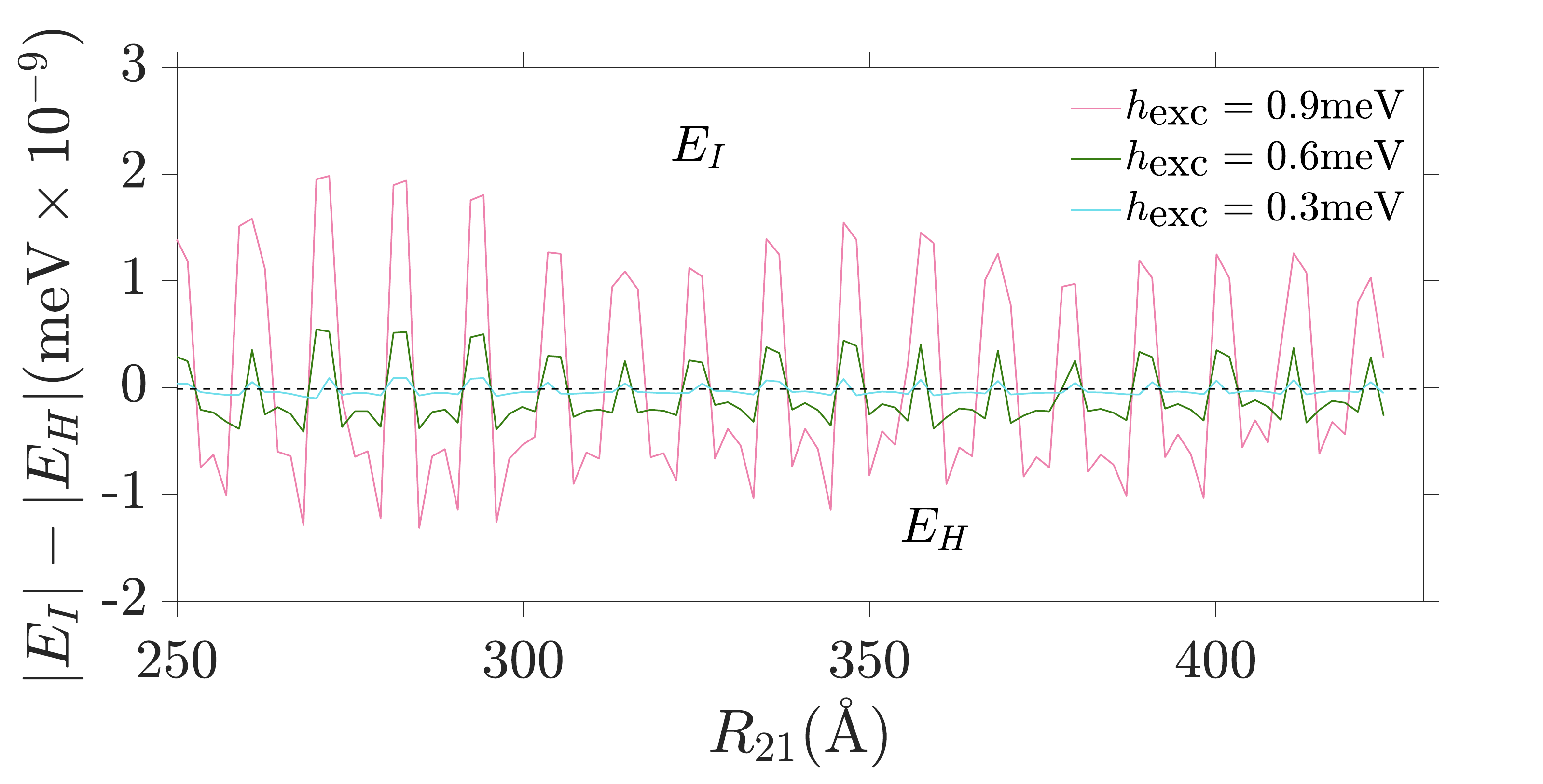}%
}
\caption{\small Difference between the Ising and Heisenberg energies for three different external values of the spin-splitting field at $T = 4$ K for the superconducting state.}
\label{fig:T_4K_vs_R21_E_I_minus_E_H}
\end{figure}

Moreover, Fig. 5 shows that the interaction between the two impurity spins still oscillates between Heisenberg and Ising terms as a function of the distance between the two impurity spins even for the case of higher temperatures $T \lesssim T_c$. The magnitude of the oscillations in Fig.\! \ref{fig:T_4K_vs_R21_E_I_minus_E_H} increases with $h_\text{exc}$ in both cases. This is reasonable since the spin-rotational invariance becomes more strongly broken with increasing $h_\text{exc}$, making the Ising and Heisenberg configurations more distinct in energy.

\subsubsection{Discussion of experimental aspects}
\begin{figure}[H]
\centering
\includegraphics[width=0.8\columnwidth,trim= 0.2cm 3.0cm 0.1cm 3.0cm,clip=true]{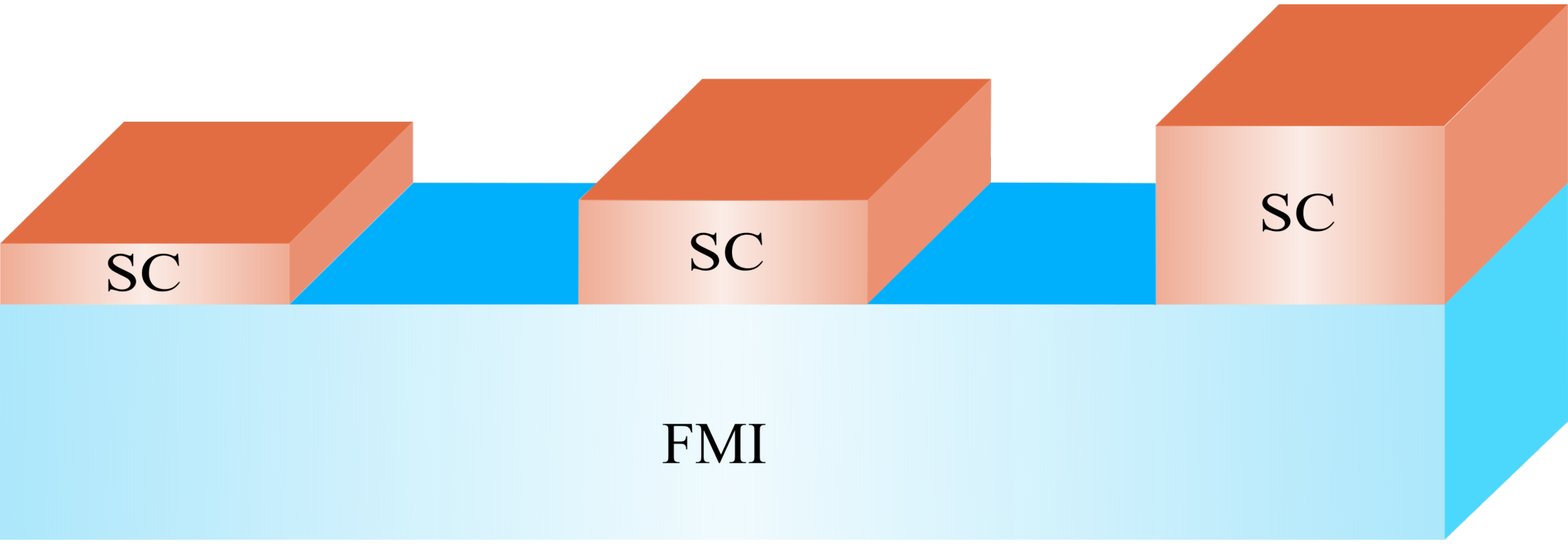}%
\caption{\small Possible experimental setup that can be used to test the effect on the RKKY energies when changing the effective Zeeman-splitting in the superconductor. By growing several superconducting layers on top of a ferromagnetic insulators and making the thickness of each superconducting layer different, the effective spin-splitting experienced by magnetic impurities placed on top of the superconducting surfaces will be different. The thickness of the superconducting layers should in all cases be much smaller than the penetration depth $\lambda$ and smaller than the superconducting coherence length $\xi_S$ in order to justify the approximation of a homogeneous spin-splitting field. }
\label{fig:model_2}
\end{figure}

We close this section by discussing possible experimental realizations of the proposed system. The magnitude of the spin-splitting field $h_\text{exc}$ can be readily tuned by an external magnetic field. Alternatively, the spin-splitting can be induced by proximity coupling the superconductor to a ferromagnetic insulator (FMI), as displayed in Fig.\! \ref{fig:model_2}. An effective  spin-splitting field in the superconductor then arises from quasiparticle reflections at the interface between the superconductor and the ferromagnet. The spin-splitting field can be assumed to be uniform if the thickness of the superconductor is much smaller than the coherence length. Also, the magnitude of the spin-splitting scales as one over the thickness of the superconducting layer \cite{bergeret_rmp_17}. The effective exchange field in the superconductor $h_{\text{exc}}$ can therefore be tuned through the thickness of the superconducting layer. Fig.\! \ref{fig:model_2} illustrates such a set up where several superconducting samples with varying thickness are grown on top of the same FMI layer. Magnetic impurity spins placed on the top surface of the superconductor will then couple via quasiparticles that experience different values of the effective $h_\text{exc}$, depending on the thickness of the superconducting layer.\\
\indent For RKKY interaction in spin-polarized systems \cite{RKKY_Ising_1}, an important point to note is that the preferred direction of the impurity spins will not be solely determined by the RKKY interaction. There are also local effective anisotropy terms of the type $E_z (S_j^z)^2$ and $E_{xy} [(S_j^x)^2 + (S_j^y)^2]$ for both impurities $j=1,2$ that are contained in $E_0$ in Eq. (\ref{eq:e0}). Moreover, when inducing a magnetization in the superconductor, there will be a coupling between the induced magnetization and the impurities, which is first order in the perturbation parameter $J$ and therefore able to dominate over the RKKY interaction for sufficiently large spin-splitting. As the interaction between the impurity spins and the homogeneous magnetization of the superconductor will be equal for both impurities, this interaction will act to align the impurity spins. If the spin-splitting arises from an external magnetic field, there will in addition be a direct Zeeman coupling to the impurity spins. This direct Zeeman coupling, which would otherwise typically be the dominant interaction determining the impurity spin orientation, can be avoided by inducing the spin-splitting through proximity to a ferromagnet.\\
\indent We want to underline that, although there will be other interactions influencing the magnetic impurity configuration, the RKKY interaction is detectable in experiments as it is the only interaction that depends on the relative orientation of the impurity spins and the distance between them. A possible experiment probing the RKKY interaction could be as follows. Consider the setup in Fig.\! \ref{fig:model_2}. The impurity spins in the superconductor will prefer to align due to the coupling to the exchange field. Using e.g. spin-polarized scanning tunneling microscopy, the energy needed to flip one of the two spins can be measured \cite{STM_experiment_2010, STM_theory_2010}. The energy necessary to flip this spin at a given impurity separation distance will be decided by the RKKY interaction as well as other present interactions. By subtracting the energy necessary to flip a spin in the absence of RKKY interaction (when there is no other impurity nearby), the RKKY interaction can then be determined.

\section{Summary}\label{summary}
In conclusion, we have determined the RKKY interaction between magnetic impurities in a spin-split superconductor, in which case the interaction becomes anisotropic in spin space. The magnitudes of the Ising and Heisenberg terms of the RKKY interaction alternate on being the dominant term and oscillate as a function of distance between the impurities, both at low temperatures $T\ll T_c$ and high temperatures $T \lesssim T_c$. 

We also demonstrate that it is possible to change the preferred orientation of the RKKY interaction from an antiferromagnetic configuration of impurity spins to a parallel configuration by adjusting the magnitude of the spin-splitting field $h_\text{exc}$. Such an effect is in principle also attainable in the normal-state of the system, but the effect is considerably more robust in the superconducting state where it occurs for a much larger set of separation distances between the impurities compared to the normal-state.

\section{Acknowledgements}
We thank E. Erlandsen for fruitful discussions. This work was supported by the Research Council of Norway through its Centres of Excellence funding scheme grant 262633.

\appendix 
\section{Bogoliubov-de Gennes transformation}
\label{Appendix_A}

In this section, we give a brief derivation of Bogoliubov-de Gennes transformation in Eq.\! \eqref{BdG_transformation}. We first rewrite Eq.\! \eqref{Hamiltonian_k_1} as follows,

\begin{align} \label{Hamiltonian_k_2}
    \begin{aligned}
   H_0 = & \frac{1}{2} \sum_{\bm{k},\sigma}
  \left( {\begin{array}{cc}
   c_{\bm{k},\sigma}^{\dagger} & c_{-\bm{k},-\sigma}
    \end{array} } \right)
    \left( {\begin{array}{cc}
   \zeta_{\bm{k}} - \sigma h_{\text{exc}} & -\sigma \Delta\\
   -\sigma \Delta & -\zeta_{\bm{k}} - \sigma h_{\text{exc}}
    \end{array} } \right) \times \\\\
  & \left( {\begin{array}{c}
    c_{\bm{k},\sigma} \\
    c_{-\bm{k},-\sigma}^{\dagger}\\
  \end{array} } \right) - \frac{|\Delta|^2}{V} + \sum_{\bm{k}} \zeta_{\bm{k}} = \frac{1}{2} \sum_{\bm{k},\sigma} \varphi_{\bm{k},\sigma}^{\dagger} M \varphi_{\bm{k},\sigma} \\
  & - \frac{|\Delta|^2}{V} + \sum_{\bm{k}} \zeta_{\bm{k}} .
    \end{aligned}
\end{align} 
\noindent In order to diagonalize the Hamiltonian, we consider the unitary matrix $P_{\bm{k},\sigma}$ of the form

\begin{align} \label{eq:eq15}
    \begin{aligned}
        & P_{\bm{k},\sigma}^{\dagger} = (\Phi_{\bm{k},\sigma}^+ \Phi_{\bm{k},\sigma}^- ),
\\
        &\Phi^+=
         \left( {\begin{array}{c}
         \upsilon_{\bm{k}}\\
         - \sigma \nu_{\bm{k}}
         \end{array} } \right), 
        \Phi^-=
        \left( {\begin{array}{c}
         \sigma \nu_{\bm{k}}\\
         \upsilon_{\bm{k}}
         \end{array} } \right),
    \end{aligned}
\end{align}
\noindent where $\Phi^+$ and $\Phi^-$ are eigenvectors of $M$. The Hamiltonian then takes the form

\begin{equation}\label{eq:eq11}
  H_0 = \frac{1}{2} \sum_{\bm{k},\sigma} \tilde{\varphi}_{\bm{k},\sigma}^{\dagger} \tilde{M} \tilde{\varphi}_{\bm{k},\sigma} - \frac{|\Delta|^2}{V} + \sum_{\bm{k}} \zeta_{\bm{k}} .
\end{equation}
\noindent We have used

\begin{align} \label{eq:eq12}
    \begin{aligned}
        & \tilde{M}=
         \left( {\begin{array}{cc}
          E_{\bm{k}, \sigma}^{+} & 0\\
          0 & E_{\bm{k}, \sigma}^{-}
          \end{array} } \right), \\
         & \tilde{\varphi}_{\bm{k},\sigma} = P_{\bm{k},\sigma} \varphi_{\bm{k},\sigma} =
         \left( {\begin{array}{c}
          \gamma_{\bm{k} , \sigma}  \\
          \gamma_{-\bm{k} , -\sigma}^{\dagger} 
    \end{array} } \right). 
    \end{aligned}
\end{align}
\noindent Here, the quasiparticle energies are $E_{\bm{k}, \sigma}^{\pm}= \pm \sqrt{\zeta_{\bm{k}}^2 + (-\sigma \Delta)^2} - \sigma h_{\text{exc}}$. Using $P_{\bm{k},\sigma}^{\dagger} \tilde{\varphi}_{\bm{k},\sigma} = \varphi_{\bm{k},\sigma}$ leads to the transformation between normal creation and annihilation operators and quasiparticle creation and annihilation operators (Eq.\! \eqref{BdG_transformation}).

\section{Effective Hamiltonian}
\label{Apendix_B}

In order to obtain the Ising and Heisenberg terms of the RKKY interaction, we calculate the expectation value of the effective Hamiltonian following the procedure outlined in section~\ref{model}. We then obtain

\begin{align} \label{<H_eff>}
    \begin{aligned}
        & \langle \Tilde{H} \rangle = \sum_{\bm{k}, \sigma} E_{\bm{k}, \sigma} n(E_{\bm{k}, \sigma}) - \frac{1}{2} \sum_{\bm{k},\bm{k^{\prime}} \atop \alpha, \beta} \sum_{i,j} (\frac{J}{N})^2 e^{i(\bm{k^{\prime}}-\bm{k}) \cdot (\bm{r}_j - \bm{r}_i)} 
        \\
         & \times \Big[ |\upsilon_{\bm{k}} \upsilon_{\bm{k^{\prime}}}|^2 \frac{n(E_{\bm{k}, \alpha})-n(E_{\bm{k^{\prime}} , \beta})}{E_{\bm{k^{\prime}} , \beta}-E_{\bm{k}, \alpha}} \bm{S}_i^{\alpha \beta} \bm{S}_j^{\beta \alpha} 
         \\
        & + \alpha \beta \upsilon_{\bm{k}}^{\ast} \upsilon_{\bm{k^{\prime}}} \nu_{-\bm{k}}^{\ast} \nu_{-\bm{k^{\prime}}} \frac{n(E_{\bm{k^{\prime}}, \beta})-n(E_{\bm{k} , \alpha})}{E_{\bm{k^{\prime}} , \beta}-E_{\bm{k}, \alpha}} \bm{S}_i^{\alpha \beta} \bm{S}_j^{-\alpha, -\beta}  
         \\
         & + (-\alpha \beta) \upsilon_{\bm{k}}^{\ast} \upsilon_{-\bm{k^{\prime}}} \nu_{\bm{k^{\prime}}} \nu_{-\bm{k}}^{\ast} \frac{n(E_{\bm{k}, \alpha})+n(E_{-\bm{k^{\prime}} , -\beta})-1}{E_{\bm{-k^{\prime}} , -\beta}+E_{\bm{k}, \alpha}} \bm{S}_i^{\alpha \beta} \bm{S}_j^{-\alpha, -\beta} 
         \\
         & + \upsilon_{\bm{k}}^{\ast} \upsilon_{\bm{k}} \nu_{\bm{k^{\prime}}} \nu_{\bm{k^{\prime}}}^{\ast} \frac{-n(E_{\bm{k}, \alpha})-n(E_{-\bm{k^{\prime}} , -\beta})+1}{E_{\bm{k} , \alpha}+E_{-\bm{k^{\prime}}, -\beta}} \bm{S}_i^{\alpha \beta} \bm{S}_j^{\beta \alpha} 
         \\
         & - \upsilon_{\bm{k^{\prime}}} \upsilon_{\bm{k^{\prime}}}^{\ast} \nu_{\bm{k}}^{\ast} \nu_{\bm{k}} \frac{n(E_{-\bm{k}, -\alpha})+n(E_{\bm{k^{\prime}} , \beta})-1}{E_{\bm{k^{\prime}} , \beta}+E_{-\bm{k}, -\alpha}} \bm{S}_i^{\alpha \beta} \bm{S}_j^{\beta \alpha} 
         \\
         & - (-\beta \alpha) \upsilon_{\bm{k^{\prime}}} \upsilon_{-\bm{k}}^{\ast} \nu_{\bm{k}}^{\ast} \nu_{-\bm{k^{\prime}}} \frac{-n(E_{-\bm{k}, -\alpha})-n(E_{\bm{k^{\prime}} , \beta})+1}{E_{\bm{k^{\prime}} , \beta}+E_{-\bm{k}, -\alpha}} \bm{S}_i^{\alpha \beta} \bm{S}_j^{-\alpha, -\beta} 
         \\
         & + (\beta \alpha) \upsilon_{-\bm{k}}^{\ast} \upsilon_{-\bm{k^{\prime}}} \nu_{\bm{k}}^{\ast} \nu_{\bm{k^{\prime}}} \frac{n(E_{-\bm{k}, -\alpha})-n(E_{-\bm{k^{\prime}} , -\beta})}{E_{-\bm{k} , -\alpha}-E_{-\bm{k^{\prime}}, -\beta}} \bm{S}_i^{\alpha \beta} \bm{S}_j^{-\alpha, -\beta}
         \\
        & + | \nu_{\bm{k}} \nu_{\bm{k^{\prime}}}|^2 \frac{n(E_{-\bm{k^{\prime}}, -\beta})-n(E_{-\bm{k} , -\alpha})}{E_{-\bm{k} , -\alpha}-E_{-\bm{k^{\prime}}, -\beta}} \bm{S}_i^{\alpha \beta} \bm{S}_j^{\beta \alpha} \Big].
    \end{aligned}
\end{align}
Here, we have defined $\bm{S}_i^{\alpha \beta} = \bm{S}_i \cdot \bm{\sigma}_{\alpha \beta}$. The first term is a constant that is not relevant for the RKKY interaction. Performing the Pauli matrix products, the second term in Eq.\! ~\eqref{<H_eff>} leads to the RKKY interaction presented in Eqs.\! ~\eqref{E_I} and ~\eqref{E_H}.
 

\bibliographystyle{apsrev4-1}
\addcontentsline{toc}{chapter}{\bibname}
\bibliography{Refs}

\end{document}